\documentclass[english,aps, prl, twocolumn]{revtex4}
\usepackage{amsmath}
\usepackage{graphicx}

\makeatletter

\newcommand{\lyxdot}{.}

\makeatother

\usepackage{babel}
\begin{document}

\title{Is transport in time-dependent random potentials universal ?}

\author{Yevgeny Krivolapov}

\affiliation{Physics Department, Technion - Israel Institute of Technology, Haifa
32000, Israel.}

\author{Shmuel Fishman}

\affiliation{Physics Department, Technion - Israel Institute of Technology, Haifa
32000, Israel.}
\begin{abstract}
The growth of the average kinetic energy of classical particles is
studied for potentials that are random both in space and time. Such
potentials are relevant for recent experiments in optics and in atom
optics. It is found that for small velocities uniform acceleration
takes place, and at a later stage fluctuations of the potential are
encountered, resulting in a regime of anomalous diffusion. This regime
was studied in the framework of the Fokker-Planck approximation. The
diffusion coefficient in velocity was expressed in terms of the average
power spectral density, which is the Fourier transform of the potential
correlation function. This enabled to establish a scaling form for
the Fokker-Planck equation and to compute the large and small velocity
limits of the diffusion coefficient. A classification of the random
potentials into universality classes, characterized by the form of
the diffusion coefficient in the limit of large and small velocity,
was performed. It was shown that one dimensional systems exhibit a
large variety of novel universality classes, contrary to systems in
higher dimensions, where only one universality class is possible.
The relation to Chirikov resonances, that are central in the theory
of Chaos, was demonstrated. The general theory was applied and numerically
tested for specific physically relevant examples.
\end{abstract}
\maketitle
Dynamics in potentials which are random both in space and time were
subject of many sophisticated studies for nearly 100 years \cite{Langevin1908,Uhlenbeck1930,Sturrock1966,Kampen2007}.
The response to forces resulting of such potentials typically differs
from ordinary diffusion. Specifically, the diffusion coefficients
predicted by such mechanisms sensitively depend on the velocity of
the particles. Also, if the potential is time-dependent, the energies
of the particles will not be constant. The existence of such `anomalous
diffusion' has been demonstrated for classical dynamics with spatially
and temporally fluctuating potentials \cite{Golubovic1991,Rosenbluth1992,Arvedson2006,Bezuglyy2006,Aguer2009,Bezuglyy2012}.
These works claim universal behavior in the sense that for generic
random potentials the diffusion coefficient exhibits a universal power-law
dependence on instantaneous velocity $v$, such that $D(v)\sim|v|^{-3}$
as $|v|\to\infty$. This in turn implies that asymptotically in time
the average velocity satisfies $\left\langle v^{2}\right\rangle \sim t^{2/5}$.
In the present Letter we will demonstrate that this picture should
be revised, and we will introduce such a revision. The potentials
we study here are continuous in space and are fundamentally different
from lattice models where the velocity (momentum) is inherently bounded.

In the present work we consider the classical dynamics of a particle
in potentials that are random both in space and in time, emphasizing
the spreading of the velocity acquired by the particle, as time evolves.
This is a fundamental problem, which was motivated by experiments
in optics \cite{Schwartz2007} and in atom optics \cite{Lye2005,Sanchez-Palencia2007},
, where the random potential is introduced by transforming an intensity
pattern into an effective potential, for the light \cite{Efremidis2002,Fleischer2003}
or the cold atoms \cite{Lye2005,Sanchez-Palencia2007}. Such potentials
are naturally described in terms of the Fourier spectrum of the waves
inducing them, and their spectral coefficients are assumed to be independent
random variables. In addition to the fundamental interest, the calculations
of the present work are also relevant for spreading of waves in random
time-dependent potentials, because at least in the regime of large
velocities (short wave-length) it is generally believed that a classical
picture is appropriate. Therefore, we may expect that the classical
results, which are presented in this Letter are relevant also for
the wave experiments (e.g., \cite{Lye2005,Schwartz2007,Sanchez-Palencia2007})
. We will consider a general stationary (both in space and time) random
potential, which is also isotropic on average. Such potentials are
conveniently described by their Fourier components,
\begin{equation}
V^{\left(1\right)}\left(\mathbf{x},t\right)=\int\hat{V}\left(\mathbf{k},\omega\right)\exp i\left(\mathbf{k}\cdot\mathbf{x}-\omega t\right)\mathrm{d}\mathbf{k}\mathrm{d}\omega+c.c,\label{eq:1}
\end{equation}
where $\hat{V}\left(\mathbf{k},\omega\right)$ is a random field chosen
such that the distribution of $V^{\left(1\right)}\left(\mathbf{x},t\right)$
is stationary both in time and space, in such case $\left\langle \hat{V}\left(\mathbf{k},\omega\right)\hat{V}\left(\mathbf{k}',\omega'\right)\right\rangle =V_{0}^{2}\delta\left(\mathbf{k}-\mathbf{k}'\right)\delta\left(\omega-\omega'\right)$,
where $\left\langle .\right\rangle $ denotes the ensemble average.
For time dependent potentials the velocity of the particles may grow,
since energy is not conserved. The dominant mechanism of the growth
of the velocity is via Chirikov resonances \cite{Zaslavskiii1972},
namely, resonances between the particle dynamics and the external
driving. Those resonances occur when the phases in (\ref{eq:1}) are
stationary. For a given velocity $\mathbf{v}=\dot{\mathbf{x}},$ this
happens for
\begin{equation}
\mathbf{k}\cdot\mathbf{v}=\omega.\label{eq:Chirikov_cond}
\end{equation}
In one dimensional systems this condition reduces to $kv=\omega$,
namely, when the particle just `surfs' on one of the waves composing
(\ref{eq:1}). The growth of the velocity occurs only when the particle
`jumps' to a wave traveling with a nearby and larger velocity than
the velocity of its initial carrier. This provides an intuitive picture
for the understanding of transport in phase-space, since the growth
of the velocity is limited to regions in phase-space where the density
of Chirikov resonances is non-zero. In dimensions two and higher particles
may increase their velocity beyond the velocity of the wave by traveling
in a direction that is not parallel or even perpendicular to the wave
propagation direction. This essential difference will yield distinct
much richer transport behavior in one dimensional systems than in
higher dimensions, as will be explained in what follows \cite{Krivolapov2012}.

Since, in this work we consider only potentials continuous in both
time and space it is natural to define the characteristic length and
time scales of the potential, $l_{x}$ and $l_{t}$, correspondingly.
Those, scales are defined such that the variation of the potential
over them is limited, and the particle will encounter an almost constant
force. Note, that $l_{t}$ and $l_{x}$ are generally different from
the exponential decay rates of the correlation function of the potential.
If the force is weak in the sense, $F<l_{x}/l_{t}^{2}$, particles
with velocities $v<l_{x}/l_{t}$ (as will be assumed in the present
work) will experience an uniform acceleration up to time $t<l_{t}$.
For longer time-scales the variations of the potential become apparent
for the particle, and the velocity will exhibit anomalous diffusion.

To calculate the anomalous diffusion coefficient we invoke the Fokker-Planck
approximation, assuming that the force is sufficiently weak and the
decay of the potential correlations, $C\left(\mathbf{x}_{1}-\mathbf{x}_{2},t_{1}-t_{2}\right)=\left\langle V\left(\mathbf{x}_{1},t_{1}\right)V\left(\mathbf{x}_{2},t_{2}\right)\right\rangle ,$
is sufficiently rapid, so that the velocity can be considered constant
on the time scale where the correlation function is appreciable. For
stationary random potentials it is convenient to use the average power
spectral density (PSD), $S\left(\mathbf{k},\omega\right)$, defined
as the Fourier transform of the correlation function (Weiner-Khinchin
theorem),
\begin{equation}
C\left(\mathbf{x},t\right)=\int\mathrm{d}\omega\int\mathrm{d}\mathbf{k}\, S\left(\mathbf{k},\omega\right)\exp i\left(\mathbf{k}\cdot\mathbf{x}-\omega t\right).\label{eq:Weiner-Khintchin}
\end{equation}
The benefit of this representation is twofold. First, in experiments
\cite{Lye2005,Sanchez-Palencia2007,Schwartz2007,Levi2011,Levi2011a}
the PSD is the naturally controlled rather than the correlation function.
Second, the analytical relations obtained in this representation are
more transparent. For a potential which is not only stationary, but
also has an isotropic PSD, and initial distribution which is isotropic
as well, the Fokker-Planck equation for the velocity is effectively
one dimensional \cite{Kampen2007},
\begin{equation}
\frac{\partial P}{\partial t}=\left(v^{-\left(d-1\right)}\frac{\partial}{\partial v}v^{d-1}D\left(v\right)\frac{\partial}{\partial v}\right)P,\label{eq:FP_mom_asymp}
\end{equation}
where $P\left(\mathbf{v},t\right)$ is the probability density and
$D\left(v\right)$ is the diffusion coefficient given by
\begin{equation}
D\left(v\right)=\pi\int\mathrm{d}\mathbf{k}\,\left(\mathbf{k}\cdot\hat{v}\right)^{2}S\left(\mathbf{k},\mathbf{k}\cdot\mathbf{v}\right).\label{eq:Dv_d}
\end{equation}
Note, that if the PSD has some typical scales, $k_{0}$ and $\omega_{0}$
then the Fokker-Planck equation is invariant under the transformation
of variables,
\begin{eqnarray}
\mathbf{v} & \to & \mathbf{v}'\nonumber \\
t & \to & t'\left(\pi V_{0}^{2}\frac{k_{0}^{2}}{\omega_{0}}\right)^{-1},
\end{eqnarray}
which allows the rescaling of the Fokker-Planck equation to a universal
form (where $V_{0}$ is the amplitude of the potential). For dimensions
two and higher, changing variables to $y=v\cos\theta$ and expanding
the angular part of the the integrand in (\ref{eq:Dv_d}) gives the
asymptotic behavior,
\begin{equation}
D\left(v\right)\sim\frac{D_{3}}{v^{3}},\label{eq:Dv_uni}
\end{equation}
 with
\begin{equation}
D_{3}=2S_{d}\int_{0}^{\infty}\mathrm{d}y\int_{0}^{\infty}\mathrm{d}k\, y^{2}k^{d+1}S\left(k,ky\right),\label{eq:D0_univ}
\end{equation}
where $S_{d}$ is the surface of a $d-$dimensional hyper-sphere.
Using this asymptotics an asymptotic scaling solution to the Fokker-Planck
equation may be obtained, $P\left(v,t\right)=t^{-d/5}g\left(v^{5}/t\right)$,
which yields the growth of the mean kinetic energy as, $\frac{1}{2}\left\langle v^{2}\right\rangle \sim t^{2/5}.$
This behavior is considered in the literature as universal \cite{Golubovic1991,Arvedson2006,Bezuglyy2006,Bezuglyy2012}
for any dimension given that the correlation function of the potential
is sufficiently differentiable. For dimensions two and higher as shown
above this is indeed the case, however for one dimensional systems
other behaviors are possible.

\begin{figure}
\begin{centering}
\includegraphics[width=8cm]{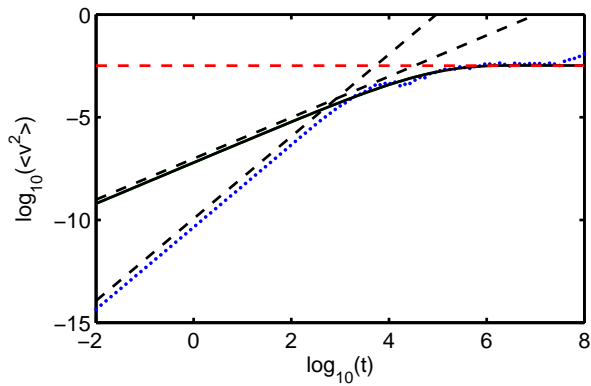}
\par\end{centering}

\caption{\label{fig:1d-opt-disk}A log-log plot of average squared velocity
as a function of time for one dimensional system with a potential
$V\left(x,t\right)=\left|U\left(x,t\right)\right|^{2}$ and a uniform
distribution of wave-numbers. The blue dots represent the result of
Monte-Carlo simulation the corresponding Newton equations averaged
over $20$ realizations and the black solid line is the numerical
solution of the Fokker-Planck equation for the velocity (\ref{eq:FP_mom_asymp}).
The dashed black and red lines are guides for the eye with the corresponding
slopes of 2 , 1 and 0. The initial condition was a narrow distribution
of velocities for the Fokker-Planck Eq. and $x=v=0$ for the Monte-Carlo
calculation. The parameters used for this simulation are, $V_{0}=10^{-4}$,
$k_{R}=0.1$.}
\end{figure}

For one dimensional systems new possibilities arise from the fact
that the angular part in (\ref{eq:Dv_d}) is missing and therefore
the diffusion coefficient,
\begin{equation}
D\left(v\right)=\pi\int k^{2}S\left(k,kv\right)\mathrm{d}k.\label{eq:Dv_1d}
\end{equation}
may acquire various asymptotic behaviors dictated by $S\left(k,\omega\right)$.
Consider a diffusion coefficient which decreases with velocity faster
than any power law, then clearly the asymptotic expansion in $v^{-1}$
is not useful, since it will produce a nil result. A simple example
is given by a PSD with the property, $S\left(k,\omega\right)=0\qquad\omega/k>v_{\max},$
which renders the diffusion coefficient zero for large velocities,
$D\left(v\right)=0\qquad v>v_{\max},$ as demonstrated in \cite{Krivolapov2012}
for a specific example. We will now demonstrate this behavior, using
an experimentally relevant potential which is proportional to the
intensity, $V\left(x,t\right)=\left|U\left(x,t\right)\right|^{2}$
of some complex field, which in turn is a superposition of waves $U\left(x,t\right)=\int\mathrm{d}k\,\hat{U}\left(k\right)\exp i\left(kx-\omega\left(k\right)t\right),$
with some dispersion relation $\omega\left(k\right)$. These potentials
appear in experiments with neutral atoms \cite{Lye2005,Sanchez-Palencia2007}
and in some experiments in optics \cite{Schwartz2007,Levi2011,Levi2011a},
where the variation of the refractive index (which plays the role
of the potential) of a photosensitive material is proportional to
the intensity of light \cite{Efremidis2002,Fleischer2003}. For simplicity
we will assume the dispersion relation to be $\omega\left(k\right)=k^{2}/2$
(which naturally appears in the framework of the paraxial approximation
in optics), and $f\left(k\right)$ to be the probability density of
the wave numbers $k$. For this type of potentials one can readily
obtain both the PSD, $S\left(k,\omega\right)=V_{0}^{2}\left|k\right|^{-1}f\left(\frac{k}{2}-\frac{\omega}{k}\right)f\left(\frac{k}{2}+\frac{\omega}{k}\right),$which
is singular for $k=0$, and the diffusion coefficient, $D\left(v\right)=4\pi V_{0}^{2}\int\left|q\right|f\left(q+v\right)f\left(q-v\right)\mathrm{d}q.$
It is clear that the decay of the diffusion coefficient with the velocity
is dictated by the decay of the probability density of the wave-numbers,
$f\left(k\right)$. An explicit expression may be obtained for example
for a uniform distribution of wave-numbers (in the interval $\left[-k_{R},k_{R}\right])$
giving \cite{Krivolapov2012},
\begin{equation}
D\left(v\right)=\begin{cases}
\frac{\pi V_{0}^{2}}{k_{R}^{2}}\left(k_{R}-\left|v\right|\right)^{2} & \left|v\right|<k_{R}\\
0 & \left|v\right|>k_{R}
\end{cases}.\label{eq:Dv_1d_segment}
\end{equation}
The resulting dynamics is demonstrated in Fig. \ref{fig:1d-opt-disk}.
The regime of a unit slope corresponds to regular diffusion and the
asymptotic regime corresponds to an absence of diffusion in velocity.

\begin{figure}
\begin{centering}
\includegraphics[width=8cm]{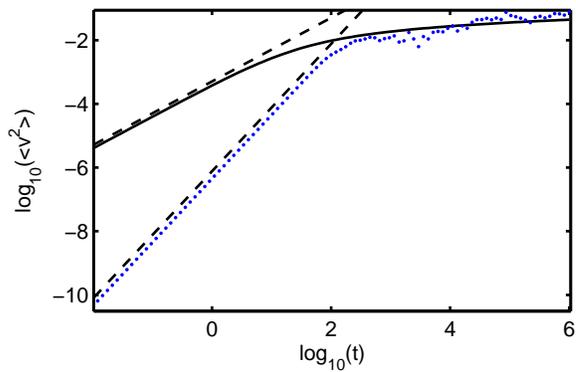}
\par\end{centering}

\caption{\label{fig:1d-opt-Gauss}Same as Fig. \ref{fig:1d-opt-disk} but with
a Gaussian distribution of wave-numbers. The dashed black lines are
guides for the eye with the corresponding slopes of 2 and 1. The parameters
used for this simulation are, $V_{0}=10^{-2}$, $k_{R}=0.1$.}
\end{figure}

Another explicit expression may be obtained for a Gaussian distribution
of wave-numbers, giving a diffusion coefficient
\begin{equation}
D\left(v\right)=2V_{0}^{2}\exp\left(-v^{2}/k_{R}^{2}\right).\label{eq:Dv_Gaussian}
\end{equation}
 The predictions of this equation are compared with Monte-Carlo simulations
and the results are presented in Fig. \ref{fig:1d-opt-Gauss}. The
correlation function for this potential can be also obtained using
(\ref{eq:Weiner-Khintchin}),
\begin{equation}
C\left(x,t\right)=V_{0}^{2}\frac{1}{\sqrt{1+k_{R}^{4}t^{2}}}\exp\left(-\frac{k_{R}^{2}x^{2}}{1+k_{R}^{4}t^{2}}\right).\label{eq:Cxt_opt_Gauss}
\end{equation}
Note, that the correlation function (\ref{eq:Cxt_opt_Gauss}) is infinitely
differentiable, decaying fast in position and slowly decaying in time.

These examples demonstrate asymptotic behavior which is very different
from (\ref{eq:Dv_uni}) described in previous studies \cite{Golubovic1991,Bezuglyy2006,Bezuglyy2012}.
Therefore new universality classes of potentials should be defined.
These universality classes are characterized by a growth in velocity
which is sub-diffusive such that the moments of the velocity distribution
grow slower than any power law. It is important to note that these
new universality classes are neither limited to potentials with a
dispersion relation nor to correlations that are slowly decaying.
For example, the PSD,
\begin{equation}
S\left(k,\omega\right)\sim\exp\left(-k_{0}/k\right)^{2}\exp\left(-\omega^{2}/\omega_{0}^{2}-k^{2}/k_{0}^{2}\right),\label{eq:Skw_essent}
\end{equation}
 gives rise to a rapidly decaying correlation function and a diffusion
coefficient which is $D\left(v\right)\sim g\left(v\right)\exp\left(-\sqrt{1+\left(v/v_{0}\right)^{2}}\right)$,
where $g\left(v\right)$ is some polynomial of $v$. The classification
of potentials into different universality classes, which differ from
(\ref{eq:Dv_uni}), may be obtained for a PSD which is not singular
and differentiable on the line $k=0$. This turns out very useful
since in some experiments the PSD could be precisely controlled, which
suggest a way to test the various claims of this Letter. The diffusion
coefficient can be expanded in powers of $v^{-1}$,
\begin{equation}
D\left(v\right)=\frac{D_{3}}{v^{3}}+\frac{D_{5}}{v^{5}}+\cdots,\label{eq:Sk_Asympt_exp}
\end{equation}
where
\begin{equation}
D_{n}=\frac{1}{\left(n-3\right)!}\int\mathrm{d}\omega\,\omega{}^{n-1}\frac{\partial^{n-3}S}{\partial k^{n-3}}\left(k,\omega\right)|_{k=0}\qquad n\geq3.
\end{equation}
If the first $n_{\max}$ derivatives, $\partial^{n}S/\partial k^{n}$,
vanish on the line $k=0$, where $n_{\max}$ is the first derivative
that is different from zero, then the resulting asymptotic behavior
is $D\left(v\right)\sim D_{n_{\max}+3}/v^{n_{\max}+3}$. The case
where \emph{all} the derivatives vanish is possible only if $S\left(k,\omega\right)$
is non-analytic on the line $k=0$. The non-analytic behavior may
be found either for a function, which is strictly zero on some finite
strip around the line $k=0$, or due to an essential singularity of
$S\left(k,\omega\right)$ on this line. The first case will lead to
a diffusion coefficient that will vanish for large velocities, for
example (\ref{eq:Dv_1d_segment}) (where the support of $f\left(k\right)$
is finite). The second case results in a decay faster than any power
law (\ref{eq:Dv_Gaussian}). By controlling the analytical behavior
of the PSD on the line $k=0$, one can vary the asymptotic behavior
of the velocity dependence of $D\left(v\right)$. The range of variation
is from $D\left(v\right)\sim v^{-3}$ , through $D\left(v\right)\sim v^{-n}$
(with $n>3$) to sub-exponential, exponential, super-exponential and
up-to $D\left(v\right)=0$ (for $v>v_{\max}$).

In this Letter the diffusion coefficient for the velocity is presented
in terms of the average power spectral density (PSD) (see Eq. (\ref{eq:Dv_d}))
for stationary potentials that are random both in space and time.
This representation is very natural for stationary (both in time and
space) potentials, in particular, for potentials which are a superposition
of waves, as they appear in optics and atom optics. The simplicity
of the expression enabled to explore the properties of the diffusion
coefficient, to establish a scaling form of the Fokker-Planck equation,
and to discover new universality classes. In particular, we were able
to calculate explicitly the diffusion coefficient for representative
examples, relevant for applications both in optics and atom optics
(Gaussian and uniform distributions of wave-vectors). We have shown
that for dimensions larger than one only one universality class is
possible in the framework of the Fokker-Planck approximation, $D\left(v\right)\sim v^{-3}$.
However for one dimensional systems new possibilities for large velocity
asymptotics were also found. It was demonstrated that diffusion in
phase-space takes place only where Chirikov resonances (\ref{eq:Chirikov_cond})
are found.

The main result of this Letter is the classification of universality
classes in one dimensional systems. In the past it was found that
in the large velocity limit the diffusion coefficient depends on the
velocity as, $D\left(v\right)\sim v^{-3}$ \cite{Rosenbluth1992,Golubovic1991,Bezuglyy2006,Bezuglyy2012}.
In the present work we have shown that this is always the case for
dimensions larger than one, and explained the mechanism of this behavior
(see discussion before (\ref{eq:Dv_uni})). However, for one dimensional
systems this is only one of the possibilities. Generally, the possible
asymptotic expansion of $D\left(v\right)$ in powers of $v^{-1}$
is:
\begin{enumerate}
\item The first term of the asymptotic expansion of $D\left(v\right)$ is
non-zero, $D\left(v\right)\sim v^{-3}$.
\item The first non-vanishing term in the asymptotic expansion of $D\left(v\right)$
is $n$ then, $D\left(v\right)\sim v^{-n}$.
\item \emph{All} terms in the asymptotic expansion of $D\left(v\right)$
\emph{are} zero, $D\left(v\right)\leq v^{-\alpha}$, for any $\alpha>0$.
In particular, the diffusion coefficient may be zero for $v>v_{\max}$
(e.g. (\ref{eq:Dv_1d_segment})) or non-zero but decreasing faster
than any power law (e.g., (\ref{eq:Dv_Gaussian})).
\end{enumerate}
All these possibilities can be realized in experiments with good control
over the PSD or more precisely the probability density of the wave-numbers,
which comprise the potential, $f\left(k\right)$. Unlike statements
of other studies \cite{Rosenbluth1992,Golubovic1991,Bezuglyy2006,Bezuglyy2012},
the differentiability of the correlation function of the potential
is not related to the classification into universality classes. Additionally,
it was demonstrated that the new universality classes do not depend
on the range of the correlation function.

For small velocities the diffusion coefficient, obtained in the framework
of the Fokker-Planck approximation, will be generally different from
zero and therefore a regime of regular diffusion is expected. However,
we have shown that initially particles will experience a uniform acceleration,
and therefore at least for short times the Fokker-Planck approximation
is invalid, as could be clearly seen from all the figures. We have
also found potentials, which produce diffusion coefficients \emph{growing}
with velocity (from zero and up to some value) \cite{Krivolapov2012}
for which the validity of the Fokker-Planck approximation is not satisfied,
since the condition that the velocity is constant during the correlation
time is violated. On a longer time-scale, the velocity will follow
a diffusion equation and eventually, it will reach the asymptotic
long-time behavior of anomalous diffusion as predicted by the Fokker-Planck
approximation.

In the present work the spreading of the velocity distribution is
studied in the framework of the Fokker-Planck equation, therefore
an obvious question to study is what happens when the Fokker-Planck
approximation fails. Another important issue, which was not addressed
in this work, is how to obtain analytically the spreading in position.
Since in the experiments that have motivated this work \cite{Schwartz2007,Levi2011a},
the relevant dynamics is of waves, rather than particles, an obvious
question to explore is the correspondence between the classical and
wave dynamics.
\begin{figure}
\begin{centering}
\includegraphics[width=8cm]{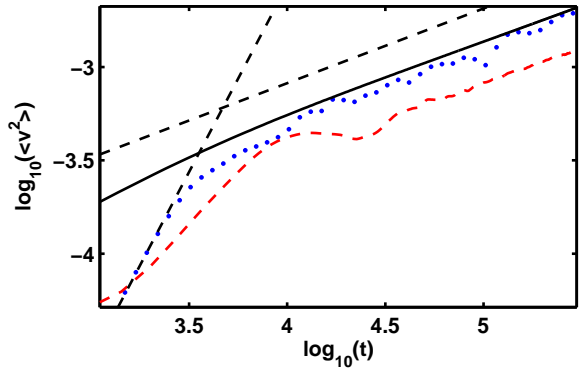}
\par\end{centering}

\caption{\label{fig:2d-opt-waves-particles}Same as Fig. \ref{fig:1d-opt-Gauss}
but for a two dimensional system, and including the wave dynamics
. The dashed black lines are guides for the eye with the corresponding
slopes of 2 and 2/5. The dashed (light) red line is the result of
a numerical simulation of the Schrödinger equation with the same potential,
averaged over 100 realizations. The initial distributions of particle
velocities are the same as the initial wavefunction, a narrow Gaussian
around the origin. The parameters used for this simulation are, $V_{0}=2.06\times10^{-4}$,
$k_{R}=0.029$ (same as experimental parameters in \cite{Levi2011a}).}
\end{figure}
This correspondence is shown in Fig. \ref{fig:2d-opt-waves-particles},
where in addition to Monte-Carlo and Fokker-Planck calculations of
the average squared velocity for particles a numerical simulation
of a corresponding wave system is presented. As is expected there
is a reasonable correspondence for large velocities. It will be explored
in further studies and the possibility for its violation is of great
interest.

This work was motivated by the experimental work of Liad Levi and
Mordechai Segev, whom we thank for many stimulating discussions and
for providing crucial insight for this problem. It is our great pleasure
to thank Tom Spencer for introducing us to \cite{Golubovic1991,Rosenbluth1992}
and to Michael Wilkinson for introducing us to \cite{Bezuglyy2006,Bezuglyy2012}.
Many of the results of the present work originated from fruitful discussions
with Michael Wilkinson during his visit to the Technion. The work
was supported in part by the US-Israel Binational Science Foundation
(BSF).

\bibliographystyle{apsrev}
\bibliography{QP-potential}

\end{document}